# Enhancement of the Hall-Lorenz number in optimally doped $YBa_2Cu_3O_{7-d}$

M. Matusiak,[1] K. Rogacki,[1,*] and B. W. Veal [2]

[1] *Institute of Low Temperature and Structure Research, Polish Academy of Sciences,*

*P.O. Box 1410, 50-950 Wrocław, Poland*

[2] *Materials Science Division, Argonne National Laboratory, Argonne, Illinois 60439, USA*

## Abstract

Electronic heat transport in the normal state of a high-quality single crystal of optimally-doped superconductor $YBa_2Cu_3O_{6.95}$ was studied by measurements of longitudinal and transverse transport coefficients. For the temperature range from 100 to 300 K, the Hall-Lorenz number ($L_{xy}$) depends weakly on temperature and is about two times larger than the Sommerfeld value of the Lorenz number $L_0 = \pi^2/3$. Our results can be interpreted using a Fermi liquid model when effects of the pseudogap that opens at the Fermi level are included. However, we find that the bipolaron model can also explain both the enhanced value and the weak temperature dependence of the Hall-Lorenz number.

---

[*] Corresponding author. Tel.: +48-71-3435021; Fax: +48-71-3441029; e-mail: K.Rogacki@int.pan.wroc.pl

Knowledge of the behavior of electrons in the normal state of high temperature superconductors (HTSC) is essential for understanding their unusual charge and heat transport properties as well as the pairing mechanism in the superconducting state. In metals, charge carriers are well described in the framework of Fermi-liquid theory, which considers well defined fermionic quasiparticles with charge $e$, spin $1/2$, and an effective mass $m^*$ to be interaction dependent. One of the basic features of the Fermi liquid is a relation between thermal and electrical conductivities known as the Wiedemann-Franz (WF) law. At given temperature the WF law can be written in the convenient dimensionless form as:

$$L \equiv \frac{\kappa_{el}}{\sigma T}\left(\frac{e}{k_B}\right)^2,$$ (1)

where $\kappa_{el}$ is the electronic thermal conductivity, $\sigma$ is the electrical conductivity, $k_B$ is Boltzmann's constant, and $L$ is the Lorenz number that in standard transport theory is equal to the Sommerfeld value $L_0 = \pi^2/3$ if the mean-free-paths for transport of charge and entropy are identical. Deviations from this law can indicate that the Fermi liquid is not a ground state of the electronic system. Violation of the WF law has been reported in some cooper-oxide superconductors, $(Pr,Ce)_2CuO_4$ [1], $(Bi,Sr)_2CuO_{6+d}$ [2,3], and underdoped and optimally-doped $YBa_2Cu_3O_{7-d}$ (YBCO) [4,5,6], indicating that the elementary excitations that carry heat in these materials at low temperatures are not fermions. On the other hand, the electronic thermal conductivity of optimally-doped $Bi_2Sr_2CaCu_2O_8$ and optimally-doped YBCO has been well described at very low temperatures by the Fermi-liquid theory with the energy spectrum of d-wave symmetry as measured by a photoemission spectroscopy [7]. Other studies of the thermal conductivity in YBCO have indicated that an unconventional d-wave or extended s-wave pairing symmetry of quasiparticles is possible in this compound [8,9].

Recently, effects on transport properties resulting from the presence of a pseudogap in the electron density of states, have been examined [10]. The pseudogap, which appears in the normal state, causes an increase in the electronic thermal conductivity and results in an enhanced $L/L_0$ ratio, as reported for $(Bi,Sr)_2CuO_{6+d}$ [2,3] and for underdoped and optimally-doped YBCO [10] and EuBCO [11] single crystals. On the other hand, a much reduced $L/L_0$ ratio has been reported for optimally-doped, untwinned YBCO crystals [5,6].

A proposed resolution to this controversy, based on bipolaron theory, has been presented [12,13]. The theory can account for the huge, one order of magnitude difference in the Hall-Lorenz number by taking into account the difference between the in-plane resistivity of twinned and detwinned single crystals. However, as discussed below, it is unclear if experimental problems might arise from the use of multiple crystals to obtain the Hall-Lorentz number.

The interpretation of thermal conductivity measurements in HTSC is complicated by the difficulty of accurately separating electron and phonon contributions. The measured total thermal conductivity ($\kappa$) is the sum of the electronic ($\kappa_{el}$) and phonon ($\kappa_{ph}$) components, $\kappa = \kappa_{el} + \kappa_{ph}$, and the contribution from each component has to be determined. In conventional metals electrons carry most of heat, so the phonon contribution to the thermal conductivity can be in many cases neglected. However, in high-$T_c$ cuprates $\kappa_{ph}$ can be the same order of magnitude as $\kappa_{el}$ (e.g. in single crystals of YBCO [14] and $Bi_2Sr_2YCu_2O_8$ [15]), or even 10 - 100 times larger (e.g. in polycrystalline samples of $REBa_2Cu_3O_7$ [16]).

Recently, Zhang et al. [5] used the thermal Hall effect to extract the electronic component from the total thermal conductivity. In zero magnetic field, the observed total thermal current is parallel to the applied temperature gradient $\nabla T$. In an external field perpendicular to $\nabla T$, a Lorenz force is generated that acts only on the electronic component of the heat current, because the phonon component is in principle unaffected by the field. Thus, the transverse $\nabla T$ refers to the heat transport by the charge carriers only.

In this paper we study the thermal Hall effect in the optimally-doped YBCO superconductor and present results on the determination of the Hall-Lorenz number ($L_{xy}$), which is regarded as a direct source of information about the electronic heat current [5]. Our studies have been performed in the normal-state of an optimally-doped $YBa_2Cu_3O_{6.95}$ single crystal and they show that $L_{xy}$ depends weakly on temperature, changing from about 8 to 5 (about 35 %) when temperature decreases from 300 to 100 K. Thus, the departure from the WF law is not so evident as reported in Ref. 5, where $L_{xy}$ drops from about 2 to 0.5 (75 %) for the similar temperature range. The value of the Hall-Lorenz number, which significantly exceeds the Sommerfeld value at room temperature, is discussed and the presence of a pseudo-gap in the electronic structure is considered as an explanation.

The single crystals of $YBa_2Cu_3O_{7-d}$ were grown in a gold crucible by a conventional self-flux growth method. These were annealed in flowing oxygen for 72 hours while cooling from 500 C to 450 C. The magnetically measured superconducting transition of these crystals had an onset of $\approx 92.5 - 93.0$ K and a width of less than $\approx 1$ K as measured with 1 Oe ac-field amplitude perpendicular to the ab-plane. The crystal selected for measurements was about 1.55 mm long, 1.35 mm wide and 0.35 mm thick (along the crystallographic $c$-axis). This crystal was aged in air and then the oxygen content was estimated to be equal to 6.95 (on the basis of the in-plane thermoelectric power value at room temperature $S_{ab}(300$ K$) \approx$ - 1 μV/K [17,18,19]). The resistively measured superconducting transition temperature was $T_c = 92.6$ K and the transition width was $\Delta T_c = 0.3$ K. The crystal was twinned, therefore the coefficients measured along the $ab$-plane should represent values averaged between the $a$ and $b$ crystallographic directions.

Electrical resistivity was measured using a four-point technique. The Hall coefficient measurements were performed by a standard method in a magnetic field of 13 T. The current and magnetic field directions were reversed several times to exclude any influence of the asymmetric position of the Hall contacts and detrimental electro-motive forces. *The longitudinal and transversal thermal conductivities were measured in a single experiment*. One edge of the crystal was glued to a heat sink, and a carbon heater was painted on the opposite edge. Using this heater a longitudinal temperature gradient ($\nabla_x T$) was generated. Due to the magnetic field applied perpendicularly to $\nabla_x T$ (and parallel to the *c* axis of the YBCO crystal) a transverse temperature gradient ($\nabla_y T$) appeared. Both gradients were measured with Chromel-Constantan thermocouples. Typically $\nabla_x T$ and $\nabla_y T$ were of the order of magnitude of 1 K/mm and 10 mK/mm, respectively. $\nabla_x T$ at $B = 0$ T was used to calculate the longitudinal thermal conductivity. We usually observed $\nabla_y T \neq 0$ at $B = 0$ T due to slightly asymmetric positions of the thermocouple junctions. This inconvenience was eliminated by varying the magnetic field between -13 and +13 T and by using, for calculations of the transversal thermal conductivity, only the slope $\Delta\big(\nabla_y T(B)\big)\big/\Delta B$. Typically, $\Delta\big(\nabla_y T(B)\big)\big/\Delta B$ was of the order of 0.1 mK/(mm·T). To obtain credible results, *B* was inverted several times with various $\nabla_x T$ at every considered temperature. Moreover, to secure from detrimental heat leaks between the sample and its surrounding, appropriate precautions were taken. Namely, long-enough thermocouple wires of a very small diameter were used ($l \approx 3$ cm, $\phi = 25$ $\mu$m), the sample space was evacuated to pressure of the order of $10^{-6}$ mbar, and the crystal was surrounded by two gold plated radiation shields. The temperature of the inner shield was kept very close ($\Delta T$ less than 0.5 K) to the sample temperature. For these conditions, the possible detrimental heat leak was estimated to be negligible.

Figure 1 shows the temperature dependencies of four transport coefficients for the *ab*-plane of the $YBa_2Cu_3O_{6.95}$ single crystal. We emphasize that the longitudinal thermal and electrical conductivities ($\kappa_{xx}$ and $\sigma_{xx}$) as well as the transverse thermal and electrical conductivities ($\kappa_{xy}$ and $\sigma_{xy}$) were obtained from the *same crystal*. The measurements were performed within two weeks, thus the properties of the crystal remained unchanged due to any possible long-term relaxation processes. Figure 1(a) shows the electrical resistivity ($\rho_{xx} = 1/\sigma_{xx}$) and the longitudinal heat conductivity in the absence of a magnetic field. The temperature dependencies of the Hall coefficient, $R_H = \sigma_{xy}/(\sigma_{xx})^2 B$, and transverse thermal conductivity, $\kappa_{xy} = \nabla_y T \cdot \kappa_{xx}/\nabla_x T$, are shown in Fig. 1(b) for the same crystal. We note that the absolute value of $\kappa_{xy}$ at 300 K and its temperature dependence, $\kappa_{xy}(T)$, are almost identical to those obtained by Zhang et al. for untwined $YBa_2Cu_3O_{6.95}$ [5]. We fitted our experimental data to the function $\kappa_{xy}/B \sim T^\alpha$ with $\alpha$ = - 1.2, while Zhang et al. obtained $\alpha$ = - 1.19. These similarities in the absolute value and the temperature dependence of $\kappa_{xy}/B$, observed for twinned (our) and untwinned crystals, reveal a rather low anisotropy of the thermal conductivity in the *ab*-plane, as expected.

The Hall-Lorenz number that is related only to the electronic thermal and charge conductivities can be calculated from the modified WF law [5]:

$$L_{xy} \equiv \frac{\kappa_{xy}}{\sigma_{xy}T}\left(\frac{e}{k_B}\right)^2. \qquad (2)$$

Within the Fermi-liquid approach the regular Lorenz number (*L*) as well as $L_{xy}$ are temperature independent and equal to the Sommerfeld value if charge carriers scatter elastically [20]. There is an acceptable approximation at low temperatures where elastic scattering of electrons by impurities is dominant, and also at high temperatures where electron-phonon scattering may be regarded as elastic. In the intermediate temperature region, i.e. in the region where our studies were performed, the inelastic electron-phonon scattering more effectively disturbs the heat current than the charge current [21], and this results in a decrease of both Lorenz numbers [5]. In such a case, different mean free paths for the transport of entropy ($l_s$) and charge ($l_e$) are expected. However, the ratio

$$a_L \equiv \frac{L^2}{L_{xy}},\qquad(3)$$

should be nearly constant since $L \sim \langle l_s \rangle / \langle l_e \rangle$ and $L_{xy} \sim \langle l_s^2 \rangle / \langle l_e^2 \rangle$ [5,20]. When inelastic scattering appears, the Hall-Lorenz number falls faster than the regular Lorenz number, but both of these parameters stay correlated. This was found to be valid in copper [5]. In untwinned $YBa_2Cu_3O_{6.95}$ single crystals a dependence on temperature, $L_{xy} \approx aT$ was observed, and was interpreted as a violation of the WF law due to predominant electron-electron scattering in the YBCO type high-$T_c$ superconductors [5].

Figure 2 shows the temperature dependence of the Hall-Lorenz number for our twinned $YBa_2Cu_3O_{6.95}$ single crystal in the normal state. The results show a large enhancement of $L_{xy}$ above the $L_0$ value and a weak temperature dependence. Here, when temperature decreases from 300 to 100 K, $L_{xy}(T)$ declines by about 35 %, a relatively small violation of the WF law. For the untwinned $YBa_2Cu_3O_{6.95}$ single crystal, $L_{xy}(T)$ declines by a comparable amount, but, because of the small magnitude of $L_{xy}(T)$, the decline is about 75 % in the same temperature range [5], extrapolating to 0 at T = 0. However, any reasonable extrapolation of the $L_{xy}(T)$ data in Fig. 2, even a linear extrapolation, gives $L_{xy}(0) \geq 3$ at $T = 0$. A possible reason for the significant differences in the $L_{xy}(T)$ measurements reported here and in Ref. 5 might be that different amounts of out-of-plane impurities existed in the two $YBa_2Cu_3O_{6.95}$ crystals used in Ref. 5, one for $\kappa_{xy}$ and the second for $\sigma_{xy}$ measurements. The possible significant influence of out-of-plane impurities on the thermal Hall conductivity has been discussed in detail in Ref. 22.

Thus it is not clear if the very different $L_{xy}$ numbers, observed for twinned and untwinned samples of optimally-doped YBCO, are real or if they may be attributable to experimental problems. Resolution of this issue will require new measurements using an untwinned crystal, with all parameter measurements acquired from the same crystal. This is an important point since the seemingly incompatible results could provide a useful test of the bipolaron model [12,13].

The bipolaron model takes into account different transport properties in the $ab$-plain of twinned and untwinned single crystals. Charge carriers are treated as slow quasi-2D bipolarons (bosons) and lighter thermally activated quasi-3D single polarons (fermions), which scatter differently depending on temperature and the direction of propagation with respect to the anisotropic crystallographic structure. Similar $\kappa_{xy}$'s are expected for both crystals, since in the $ab$-plane the thermal conductivity is dominated by the 3D polarons. At the same time, $\sigma_{xy}$'s can be substantially different because the 2D bipolarons do not contribute to $\sigma_{xy}$ in the twinned crystal, whereas they do contribute in the untwinned crystal. The model numerically fits both longitudinal and transverse kinetic coefficients providing a possible explanation of our data and apparently controversial results reported in literature (see references in Refs. 12 and 13).

As shown in Fig. 2, measurements of $L_{xy}(T)$ for an optimally-doped EuBa$_2$Cu$_3$O$_7$ (EuBCO) single crystal [11] are very similar to our measurements for optimally-doped YBCO. These data show that $L_{xy}$'s have weak temperature dependences with magnitudes about two times larger than the Sommerfeld value. Similar results were also observed for an underdoped EuBCO crystal ($T_c = 61$ K) in a reduced temperature range, 160 - 300 K [11]. (Note, however, that for the underdoped sample, $L_{xy}$ deviates markedly from this behavior at T < 160 K.)

The temperature dependence of the electronic contribution to the total thermal conductivity may be derived if we assume that the scattering of electrons is approximately elastic. Then, $\kappa_{el}(T)$ can be obtained by using an assumption $L \approx L_{xy}$, Eq. 3 and the WF law (Eq. 1). The results are shown in the inset of Fig. 2. With decreasing temperature, $\kappa_{el}(T)$ goes down and this can be interpreted as a result of a decrease in the effective concentration of charge carriers. Similarities between the behavior of $\kappa_{el}(T)$ and the temperature dependence of the Hall-concentration ($n_H = 1/(e\,R_H)$), that is shown in the same inset, support this conclusion. The decrease in the effective charge carrier concentration with declining temperature may result from the presence of the pseudogap, which appears as a rapid change of the electronic density of states (EDOS) below and above the Fermi level ($E_F$) [23,10]. Since only the states within a width of the order of a few times $k_B T$ around $E_F$ contribute to transport phenomena, we can expect that the $E_F \pm k_B T$ region narrows and thus EDOS effectively reduces with declining temperature. This results in a gradual decrease of $\kappa_{el}(T)$ and $n_H(T)$.

Experimental results obtained for transverse thermal and electrical conductivities can be well fitted with the functions derived in the framework of the marginal-Fermi-liquid (MFL) approximation [22]: $\kappa_{xy}(T) = a_h T^{-1} + b_h T^{-2}$ and $\sigma_{xy}(T) = a_c T^{-2} + b_c T^{-3}$. However, fitting parameters have the ratios $b_h/a_h \approx 0.13\, b_c/a_c$, which is substantially lower than $b_h/a_h \approx 0.73\, b_c/a_c$, obtained in Ref. 22. Thus, MFL theory correctly predicts the temperature dependence of the WF law for high-$T_c$ superconductors of the YBCO type but the $L_{xy}$ values predicted by the MFL approximation are several times lower than the values obtained in our experiments. The discrepancy between experimental and theoretical results may have relevance for farther modification or development of the MFL model.

Another effort to describe electrical and thermal transport in the normal state of HTSC was made by Coleman et al. [24], who based their calculations on a modification of Anderson's Fermi liquid model. They argued that, in cuprates, the thermal transport can be understood in terms of scattering processes that are sensitive to the charge-conjugation symmetry of the quasiparticles with the quasiparticles displaying two different transport relaxation times [25]. Coleman, et. al. obtained $\kappa_{xy} \sim T^{-2}$ and $\sigma_{xy} \sim T^{-3}$, where the temperature dependences of the transport coefficients were derived using a simplified parabolic band. Therefore $\kappa_{xy}/\sigma_{xy}T$ and thus $L_{xy}$ is constant in temperature. Our experimental measurements yield somewhat different temperature dependences: $\kappa_{xy} \sim T^{-1.2}$ and $\sigma_{xy} \sim T^{-2.7}$.

The weak temperature dependence of $L_{xy}$ for optimally doped YBCO-type HTSC's is reasonably consistent with a Fermi liquid picture. The remaining question is why the value of $L_{xy}$ is higher than the Sommerfeld value of the Lorenz number ($L_0 \approx 3.3$). Despite a possible systematic error that can be as large as 30% of the obtained value (mainly due to sample geometry uncertainties), the smallest possible $L_{xy}$ at high temperatures is definitely larger than $L_0$. Furthermore, a value of Lorenz number which exceeds the Sommerfeld value has also been derived from experimental studies of the thermal transport in other HTSC at temperatures lower than $T_c$ and in high magnetic fields [1,2]. This cannot be explained within the standard metallic approach. However, an increase of $L$ is expected if the pseudo-gap that opens at the Fermi surface is taken into account, as has been shown by Minami et al. [10]. They considered a triangular pseudo-gap and found that the ratio $L / L_0$ can reach a value of $\approx 2.5$, depending on temperature and width of the pseudo-gap. This agrees very well with our results.

As mentioned above, a similarly enhanced $L_{xy}$ was also reported for underdoped and optimally-doped EuBCO crystals [11]. The experimental results were again interpreted using a phenomenological model which follows the approach of Minami et al. [10] but, for the underdoped sample, two pseudogaps, that open at the Fermi level, are considered. The appearance of two distinct pseudogaps has been reported in many experiments [see Refs in Ref. 11]. Inclusion of the pseudogaps enables the $L_{xy}(T)$ dependence observed for the underdoped EuBCO crystal to be quantitatively interpreted. These results may be viewed as additional confirmation that the pseudogap, which has been reported for most HTSC's, including recently discovered Fe pnictides [26], is an important feature responsible for some unusual electronic properties of these materials. Thus it appears that, when the pseudogap is considered, a Fermi liquid picture will suffice to account for the magnitude and temperature dependence of the Hall-Lorentz number.

In summary, we have investigated transport properties in the normal state of a superconducting $YBa_2Cu_3O_{6.95}$ single crystal. We have studied the temperature dependencies of the longitudinal electrical and thermal conductivities, as well as the Hall and thermal Hall effects represented by the transverse electrical and thermal conductivities, respectively. The obtained results have been used to calculate the temperature dependence of the Hall-Lorenz number ($L_{xy}$). This number varies slowly with temperature and is about two times larger than the Sommerfeld value of the Lorenz number. We have concluded that the normal state of optimally-doped twinned $YBa_2Cu_3O_{6.95}$ may be understood within the Fermi-liquid approximation when modified by the presence of a pseudo-gap at the Fermi level. Our results might also be interpreted using the bipolaron model, where mobile lattice bipolarons and thermally activated polarons are considered as charge carriers responsible for the normal-state transport properties of the YBCO type superconductors. However, experimental uncertainties for untwinned single crystals remain to be resolved.

**Acknowledgements**

The authors are grateful to Dr. C. Sułkowski for thermopower measurements and Dr. T. Plackowski for fruitful discussion. Support by the U. S. Dept. of Energy, Office of Science, under contract number DE-AC02-06CH11357 is acknowledged (BWV).

**Figure captions**

1. The temperature dependences of the transport coefficients for the $YBa_2Cu_3O_{6.95}$ single crystal. In the upper panel (*a*) the longitudinal electrical resistivity (solid line and left axis) and the longitudinal thermal conductivity (open points and right axis) are presented; the dashed line shows the linear fit of the resistivity in a range from 100 to 300 K. The bottom panel (*b*) shows the Hall coefficient (solid line and left axis) and the transverse thermal conductivity divided by the magnetic field (open points and right axis); the dashed line shows a function $aT^{-1.2}$ being the best power fit to the experimental points.

2. The temperature dependence of the Hall-Lorenz number (closed points) for the $YBa_2Cu_3O_{6.95}$ single crystal; the dashed line is a guide for the eye. For comparison, values of $L_{xy}$ (open diamonds) obtained for an optimally-doped $EuBa_2Cu_3O_{7-\delta}$ single crystal are also depicted [11]. The inset shows the temperature dependences of the Hall-concentration (solid line and left axis) and the electronic thermal conductivity (open points and right axis) for the $YBa_2Cu_3O_{6.95}$ crystal.

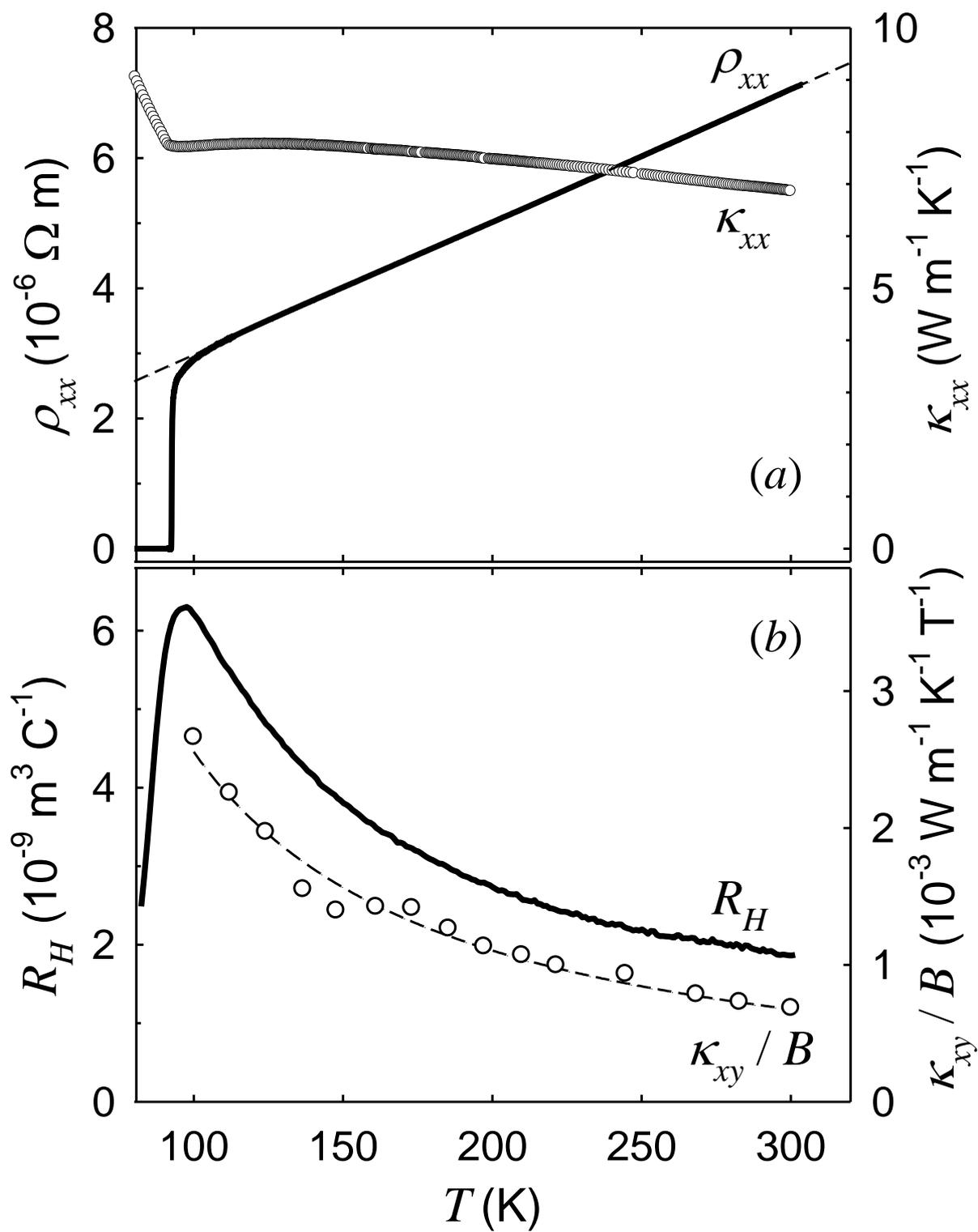

**Figure 1.**

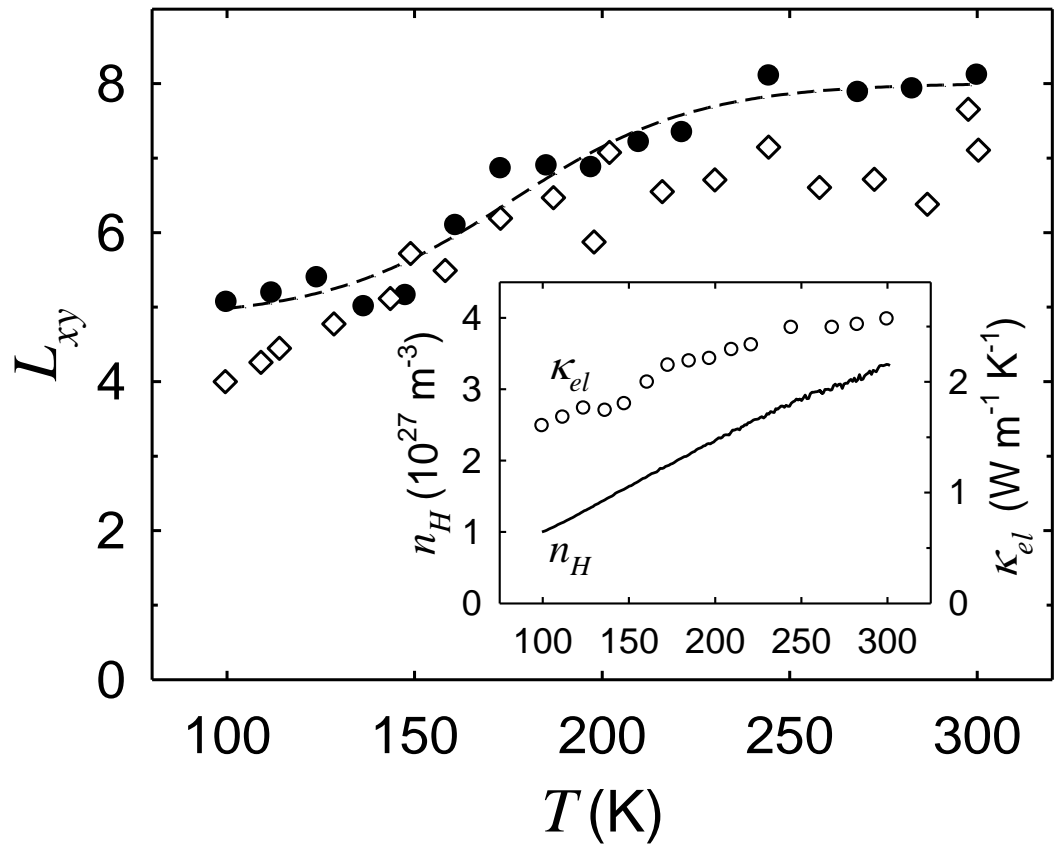

**Figure 2.**